\documentclass[%
 preprint,
showpacs,preprintnumbers,
 amsmath,amssymb,
 aps,
]{revtex4-1}

\usepackage{graphicx}
\usepackage{dcolumn}
\usepackage{bm}

\begin{document}

\preprint{LA-UR 19-26570}

\title{On the possibility of Bose Einstein condensation in lower dimensions in the thermodynamic limit}

\author{Shirish M. Chitanvis}
 \altaffiliation[Also at ]{Department of Physics and Nuclear Engineering, U.S. Military Academy at West Point, N.Y.}
 \email{shirish@lanl.gov}
\affiliation{%
 Los Alamos National Laboratory\\
 Los Alamos, N.M. 87545\\
}%




\date{\today}

\begin{abstract}
Standard arguments state that Bose Einstein condensation (BEC) cannot occur in dimensions lower than three in the thermodynamic limit
as the expressions for the number of bosons in the excited states are unbounded.
These arguments imply that the number density is infinite, which is an extraordinary condition.
As an alternative, we explore the use of
regularization techniques to show that the number density of non-interacting bosons in the excited state is finite in two and one dimensions at low temperatures,
making it possible to have a  BEC transition in the thermodynamic limit.
We suggest creating two-dimensional optical traps of increasing sizes to test this hypothesis as the thermodynamic limit is approached.

\end{abstract}

\maketitle


\section{\label{sec:level1}Introduction}

Einstein showed that Bose's derivation of Planck's law involved indistinguishability and multiple occupancy of a given energy level.
This led naturally to the consideration of the condensation of particles that follow Bose Einstein (BE) statistics.
The arguments revolve around the expression for the number of particles in excited states ($N_e$) of a non-interacting BE system in a given dimension $d$ is given by\cite{kkr}:

\begin{equation}
N_e(d,\tau) = \int_0^\infty d\epsilon ~\frac{{\cal D}_d(\epsilon)}{\exp(\frac{\epsilon - \mu}{\tau})-1}
\label{eq1}
\end{equation}
where ${\cal D}_d(\epsilon)$ is the density of states in $d$ dimensions,
$\tau = k_B T$, with $k_B$ being Boltzmann's constant, $T$ being the absolute temperature,
and $\mu$ is the chemical potential.
The number of particles $N_0$ in the ground state is given by:

\begin{equation}
N_0 = \frac{1}{\exp(-\frac{\mu}{\tau}) -1}
\label{eq2}
\end{equation}

$N_e$ has to be much less than $N_0$ in order to BEC to occur.
But a necessary condition for that to occur is that $N_e$ has to be finite.
In general, $N_e(d)$ has the following approximate form in $d$ dimensions at low temperature:

\begin{equation}
N_e(d,\tau) \approx C_d~\tau^{\frac{d}{2}}~\int_0^\infty dx~ \frac{x^{\frac{d}{2}-1}}{exp(x)-1}
\label{eq3}
\end{equation}
where $C_d$ is a constant characteristic of $d$ dimensions:

\begin{eqnarray}
C_1 &&= \frac{L}{\pi}~ \left( \frac{2 M}{\hbar^2}\right)^{1/2} \nonumber\\
C_2 &&= \frac{L^2}{2 \pi}~ \left( \frac{2 M}{\hbar^2}\right) \nonumber\\
C_3 &&= \frac{L^3}{4 \pi^2}~ \left( \frac{2 M}{\hbar^2}\right)^{3/2}
\label{eq3a}
\end{eqnarray}

It turns out that the integral in Eqn.\ref{eq3} is finite in three dimensions, but diverges in one and two dimensions. 
The divergences arise from the low end of the energy spectrum as we approach the thermodynamic limit.
The divergence is interpreted as implying that the number of bosons in the excited states is infinite and as such can never be less than the number in the ground state.
It also implies that
the relevant number density in two and one dimensions is infinite,
a somewhat problematic condition.

Gunther et al\cite{gu74} showed that in the case that the temperature is lowered in an isobaric process while maintaining an infinite number density, BEC is possible in two dimensions in the thermodynamic limit.
Later van Druten and Ketterle,\cite{dk97} as well as Bagnato and Kleppner\cite{bk91} showed that in the presence of an external potential which mimics the effects of an optical trap, BEC is permissible in
both one and two dimensions.
The external potential has to be sufficiently weak for the theorem of Bagnato and Kleppner to work.
These investigations were motivated by the construction of optical traps of finite sizes,
and therefore did not consider the thermodynamic limit.
BEC was successfully exhibited in two dimensions by Gorlitz et al\cite{gr01} and Rychtarik et al\cite{ry04} for finite systems.

We show in this paper that the invocation of regularization techniques which render finite the relevant integrals reveals that BEC is possible
in one and two dimensions in the thermodynamic limit, in the absence of external confining potentials.
We suggest creating two-dimensional optical traps of increasing sizes to test this hypothesis as the thermodynamic limit is approached.
We realize that this is not an easy task.

Our focus in this paper is on ideal bosons which do not interact with each other.
Fischer on the other hand has paid a great deal of attention to interacting, finite systems of bosons\cite{fis02}.
We intend to study interacting bosons in the near future\cite{halp19}.

\section{\label{sec:level1}Marginality of two dimensional systems}
The integral in Eqn.\ref{eq3} may be split into two parts:

\begin{equation}
N_e(d,\tau) = C_d~\tau^{\frac{d}{2}}~\left(\int_0^\alpha dx \frac{x^{\frac{d}{2}-1}}{exp(x)-1}+\int_\alpha^{\infty} dx \frac{x^{\frac{d}{2}-1}}{exp(x)-1}\right)
\label{eq3b}
\end{equation}
where it is easy to see that the right hand side of Eqn.\ref{eq3b} is independent of the value of $\alpha$.
The logic behind the split into two integrals is to demonstrate, as done in field theory, that an additive divergent part can be identified and isolated.
The finite residue is then granted physical properties.

In two dimensions the areal number density $n_e$ is governed by the behavior of the integrand in Eqn.\ref{eq3} as 
$x\to 0$:

\begin{equation}
n_e(2,\tau) \approx C_2 \tau~\int_0^\alpha \frac{dx}{x}
\label{eq4}
\end{equation}
where $n_e(2,\tau) = N_e/L^2$ is the areal number density in two dimensions,
where $c_2 = M/(\pi \hbar^2)$,
$M$ being the mass of the BE particle and $L$ being the linear extent of the system in each dimension.
The ``infra-red" divergence is caused by attempting to go to the thermodynamic limit as the linear dimension of the system $L \to \infty$,
and the discrete sum over states is converted to an integral.
The divergence of the integral is interpreted pedagogically as an indication that the number of particles in the excited states is unbound
at all temperatures and thus prohibits BEC in two dimensions.
We find this interpretation problematic as it essentially implies an infinite areal number density of bosons ($N_e/L^2$) in two dimensions.

Notice that the divergence is merely logarithmic and one can enquire what happens if an ``infrared" cutoff is employed to render the integral finite:

\begin{equation}
n_e(2,\tau) \to n_e(2,\tau,x_c) \approx  -c_2\tau~\ln(x_c)
\label{eq5}
\end{equation}

It is reasonable to assume that the cutoff $x_c \propto L^{-2}$ in two dimensions.
The logarithmic dependence on $x_c$ implies that there is only a weak dependence on the size of the system.
The finiteness of $n_e$ now implies the existence of BEC in two dimensions.
An expression for the temperature at which BEC occurs can now written down and it depends only weakly on the cutoff $x_c$.
We suggest creating two-dimensional optical traps of increasing sizes to test this hypothesis as the thermodynamic limit is approached.

An alternate way to cast the discussion regarding the cutoff is to note that  $n_e(2,\tau,x_c)$ possesses the following scaling property:

\begin{equation}
n_e(2,\tau,\lambda x_c) = n_e(2,\tau,x_c)-c_2 \tau~\ln(\lambda)
\label{eq5p1}
\end{equation}

Now BEC in two dimensions has been investigated before by
Gunther et al\cite{gu74}, who argued in essence that the divergence of $n_e$ in two dimensions could be circumvented by considering
an isobaric process.
In this sense they suggested that BEC could occur in two dimensions.
Our strategy is to use Renormalization Group theory arguments to revisit $n_e$ in two dimensions,
and we come to qualitatively similar conclusions regarding BEC in two dimensions while allowing the two dimensional number density 
to remain finite.

Renormalization group techniques involve two concepts.
One is a set of scaling transformations under which the model remains invariant.
The other is regularization techniques to render finite relevant  integrals which diverge.
These two concepts are then combined to provide insights into higher order corrections to the theory.
At this point it may be tempting to invoke scaling arguments along the lines of those employed by Abrahams et al\cite{gof79} in the context of electron localization (1978),
given the logarithmic dependence we found in Eqn.\ref{eq5}.
But the problem of electron localization is quite different in that the conductance of two-dimensional disordered systems can only be calculated approximately 
and Renormalization group theory arguments were invoked to understand higher order effects in the thermodynamic limit.

Here our focus is on more on regularization methods, as we are considering non-interacting bosons. 
In the next section we will use a method analogous to dimensional regularization to obtain an unambiguous expression for the transition temperature,
independent of the cutoff.
It is possible that scaling analyses\cite{gof79} may be useful when interacting bosons are considered.

\section{\label{sec:level2}Dimensional regularization arguments in two dimensions}
The arguments presented above leaves us with a choice to be made for $x_c$.
One can propose a cutoff based on physical arguments.
The need to make this choice can be circumvented by appealing to dimensional regularization.
Dimensional regularization in field theory revolves around the divergence of the inverse of the gamma function.
In our case, we find that it revolves around the Riemann zeta function $\zeta(s)$.

The relevant integral we need to study has the following form:

\begin{eqnarray}
{\cal I}(s) &&= \int_0^{\infty} dx \frac{x^{s-1}}{\exp(x)-1} \nonumber\\
&&= \Gamma(s)~\zeta(s)
\label{eq6}
\end{eqnarray}
where $s=d/2$.
From Erdelyi et al, one can find the following series for $\zeta(s)$ due to Hardy:

\begin{equation}
\zeta(s) = \frac{1}{s-1} + \gamma + \sum_{n=1}^{\infty} \gamma_n (s-1)^n
\label{eq6}
\end{equation}
where $\gamma=0.5772...$,
and the coefficients $\gamma_n$ in turn are given by a series expansion.

We see that for $d=2$, when $s=1$, the finite part of $\zeta(1)$ is simply $\gamma$.
We now identify the physically relevant part of the integral for $n_e$ as being proportional to $\gamma$.
The BEC temperature in two dimensions $T_{BE}(2)$ is identified when the number of bosons 
in the excited state equals the number in the ground state $N_0$:

\begin{equation}
T_{BE}(2) = n_A~\left(\frac{\pi \hbar^2}{\gamma M k_B} \right)
\label{eq7}
\end{equation}
where $n_A = N_0/L^2$ is the areal number density and 
$k_B$ is Boltzmann's constant.
For Rubidium with an areal number density of $10^{15} m^{-2}$,
we find that $T_{BE}(2) \sim 0.24 K$.

We suggest creating two-dimensional optical traps of increasing sizes to test this hypothesis as the thermodynamic limit is approached.

\section{One dimensional systems}

As in the two-dimensional case, the integral expression for the number of excited BE particles in one dimension is divergent unless an infra-red cutoff $x_c$ is imposed.
By choosing $1>>\alpha >> x_c$, the number density of bosons in the excited state $N_e/L$ can be expressed as:

\begin{eqnarray}
n_e(1,\tau,x_c) &&\approx c_1 \tau^{1/2}~\left(\int_{x_c}^\alpha \frac{dx}{x^{3/2} (1-x/2)}+J(\alpha)\right)\nonumber\\
&&\approx c_1 \tau^{1/2}~\left(-\frac{2}{3 x_c^{3/2}} +\frac{1}{2} \ln\left(\frac{1+\sqrt{\alpha/2}}{1-\sqrt{\alpha/2} }\right) + J(\alpha)+o(\alpha)\right)\nonumber\\
J(\alpha) &&= \int_\alpha^{\infty} dx \frac{x^{-1/2}}{\exp(x)-1}
\label{eq5a}
\end{eqnarray}
where $c_1 = 2M/(\pi \hbar^2)$, and
the integral denoted by $J(\alpha)$ is finite and can be evaluated numerically if required.

We propose that the divergent term ($\propto x_c^{-3/2}$) be canceled through the invocation of a counterterm as done in field theory, leaving a finite residue as $x_c \to 0$.
The physical interpretation is that the number of excited bosons in one dimension $n_e(1,\tau)$ is finite,
implying a BEC condensation in one dimension.
Note that the resultant BEC temperature is independent on the cutoff as $x_c \to 0$.

We have been unable to construct a dimensional regularization argument in one dimension similar to the two-dimensional argument in the previous section.



\section{General Discussion}
In the case of electron localization in disordered systems it was shown\cite{gof79} that while there is a smooth transition from extended to localized states in three dimensions, all states are localized in a disordered system in one and two dimensions.
This conclusion turns out to be consistent with the Mermin-Wagner theorem which was specifically proven for the case of magnetism.
Onsager's solution of the Ising model in two dimensions and the Kosterlitz-Thouless transition are notable exceptions to this theorem.
We have argued here that a transition to a BEC state is possible in three and two and one dimensions in the thermodynamic limit.

While our arguments apply strictly speaking to non-interacting systems, we hypothesize that
interacting boson systems can be adequately described via an effective mass, so that our arguments will carry over simply.
However, the problem of interactions has to be treated rigorously in the thermodynamic limit, as done by Fischer\cite{fis02} for finite systems.

\section{Conclusions}
We have argued here that a transition to a BEC state is possible in two and one dimensions in the thermodynamic limit.
Regularization techniques were invoked to draw our conclusions.
We suggest creating two-dimensional optical traps of increasing sizes to test this hypothesis as the thermodynamic limit is approached.

\section{Acknowledgments}
I would like to acknowledge Eddy Timmermans for providing me with relevant papers.

The work described in this paper was done under the auspices of the DOE while the author was stationed at West Point in
the Department of Physics and Nuclear Engineering.

%
%


\bibliography{bibliography}
\bibliographystyle{./natbib/unsrtnat}

\end{document}